\documentclass[twocolumn,secnumarabic,amssymb, nobibnotes, aps, prl, showpacs, superscriptaddress]{revtex4-1}

\usepackage{graphicx}
\usepackage{dcolumn}
\usepackage{bm}
\usepackage{color} 
\begin{document}


\title{Temperature-driven transition from a semiconductor to a topological insulator}

\author{Steffen Wiedmann}
\email{s.wiedmann@science.ru.nl}
\affiliation{High Field Magnet Laboratory and Institute for Molecules and Materials, Radboud University,
Toernooiveld 7, 6525 ED Nijmegen, The Netherlands.}

\author{Andreas Jost}
\affiliation{High Field Magnet Laboratory and Institute for Molecules and Materials, Radboud University,
Toernooiveld 7, 6525 ED Nijmegen, The Netherlands.}

\author{Cornelius Thienel}
\affiliation{Physikalisches Institut (EP3), Universit\"{a}t W\"{u}rzburg, Am Hubland, D-97074, W\"{u}rzburg, Germany.}

\author{Christoph Br\"une}
\affiliation{Physikalisches Institut (EP3), Universit\"{a}t W\"{u}rzburg, Am Hubland, D-97074, W\"{u}rzburg, Germany.}

\author{Philipp Leubner}
\affiliation{Physikalisches Institut (EP3), Universit\"{a}t W\"{u}rzburg, Am Hubland, D-97074, W\"{u}rzburg, Germany.}

\author{Hartmut Buhmann}
\affiliation{Physikalisches Institut (EP3), Universit\"{a}t W\"{u}rzburg, Am Hubland, D-97074, W\"{u}rzburg, Germany.}

\author{Laurens W.~Molenkamp}
\affiliation{Physikalisches Institut (EP3), Universit\"{a}t W\"{u}rzburg, Am Hubland, D-97074, W\"{u}rzburg, Germany.}

\author{J.~C.~Maan}
\affiliation{High Field Magnet Laboratory and Institute for Molecules and Materials, Radboud University,
Toernooiveld 7, 6525 ED Nijmegen, The Netherlands.}

\author{Uli Zeitler}
\affiliation{High Field Magnet Laboratory and Institute for Molecules and Materials, Radboud University,
Toernooiveld 7, 6525 ED Nijmegen, The Netherlands.}

\date{\today}

\begin{abstract}
We report on a temperature-induced transition from a conventional 
semiconductor to a two-dimensional topological insulator investigated by means 
of magneto-transport experiments on HgTe/CdTe quantum well structures. 
At low temperatures, we are in the regime of the quantum spin Hall effect 
and observe an ambipolar quantized Hall resistance by tuning the Fermi 
energy through the bulk band gap. At room temperature, we find electron 
and hole conduction that can be described by a classical two-carrier model. 
Above the onset of quantized magneto-transport at low temperature, we observe 
a pronounced linear magneto-resistance that develops from a classical quadratic 
low-field magneto-resistance if electrons and holes coexist. Temperature-dependent 
bulk band structure calculations predict a transition from a conventional 
semiconductor to a topological insulator in the regime where the linear 
magneto-resistance occurs. 
\end{abstract}


\maketitle

\section{Introduction}

Narrow-gap semiconductors possess conduction bands which are strongly 
non-parabolic and spin-orbit splittings that can be even larger than 
the fundamental band gap \cite{1}. A particular interesting system is 
a type-III heterostructure composed of the semimetal HgTe and the 
wide-gap semiconductor HgCdTe with a low Hg content. These quantum well
(QW) structures with an energy gap of several meV have been experimentally 
investigated by means of optics and magneto-transport already in the late 
90s in order to obtain information about the band structure (BS) 
and the Landau level dispersion \cite{2,3}. In 2006, Bernevig \textit{et al.} 
predicted that the quantum spin Hall effect (QSHE) can be observed in 
inverted HgTe/CdTe QW structures if the layer thickness is larger than 
a critical value \cite{4}. The system is then referred to as a two-dimensional (2D) 
topological insulator (TI) \cite{5,6}. The hallmark of this new state of 
matter is a quantized longitudinal conductance when the Fermi energy is in 
the bulk band gap and transport is governed by spin-polarized counter-propagating 
edge states. This quantized conductance has been found experimentally first 
in inverted HgTe QWs \cite{7} and later on in InAs/GaSb heterostructures
\cite{8,9}. The existence of the helical states has been confirmed in 
inverted HgTe QWs by non-local measurements \cite{10} and by verifying 
their spin polarization \cite{11}.

Bulk HgTe crystallizes in zincblende structure. When the semimetal
HgTe with a negative energy gap of $E_g$=-0.3~eV is combined with
the semiconductor HgCdTe, a type-III QW is formed. The band order in
HgTe QWs with HgCdTe barriers depends strongly on the quantum confinement, 
i.e., the width $d$ of the QW. For $d<d_c$, the system is a conventional 
direct band-gap semiconductor with a $s$-type $\Gamma_6$ conduction band and 
$p$-type $\Gamma_8$ valence band. $d_c$ is the critical thickness of the QW 
where the system becomes a zero-gap semiconductor \cite{12}. Calculations 
within the 8$\times$8 \textbf{k$\cdot$p} Kane model yield $d_c$=6.3~nm for a 
QW on a Cd$_{0.96}$Zn$_{0.04}$Te substrate and $d_c$=6.7~nm on a CdTe substrate 
in the zero-temperature limit. For $d>d_c$ the BS is inverted, i.e. the $H1$-band 
is the conduction and the $E1$-band is the valence band, and the system is a 2D TI. 
Moreover, the system has an indirect band-gap. If the well width is increased 
further, the confinement energy decreases and the system exhibits semi-metallic
behavior \cite{13}. 

Apart from the variation of the QW thickness, temperature can induce a phase 
transition from a normal state ($T>T_{c}$) to a topologically non-trivial state 
of matter ($T<T_{c}$). This is caused by the strong temperature dependence of the $E1$-band, 
which is the topic of this paper. In order to provide experimental evidence for
such a temperature-driven transition, we have performed magneto-transport experiments 
on four different HgTe quantum wells, all with a width exceeding the critical 
thickness $d_c$ at $T$=0. We show that we are in the regime of the QSHE and 
observe an ambipolar quantized Hall resistance at low temperature when 
the Fermi energy is tuned through the bulk band gap. In contrast, at 
room temperature we find electron and hole conduction that can be described 
by a classical two-carrier model. In an intermediate temperature range (100~K$\leq T \leq$205~K),
where Shubnikov-de Haas oscillations and quantum Hall effect are absent, we observe 
a pronounced linear magneto-resistance (LMR) that develops from a classical quadratic 
low-field magneto-resistance (MR). Bulk band structure calculations using an 
eight-band \textbf{k$\cdot$p} model demonstrate a transition from a conventional 
semiconductor to a topological insulator in the regime where the LMR is observed.

\section{Experimental details and sample characterization at $T$=4.2~K}

We have grown inverted HgTe QWs with (001) surface orientation by molecular
beam epitaxy on a CdTe (sample S1, and S4) and on a Cd$_{0.96}$Zn$_{0.04}$Te
substrate (sample S2 and S3). Details for the samples S1-S4 are given in Table T1
in the supplemental material \cite{14}. Lithographically defined Hall-bar structures 
have been produced with the dimension ($L\times W$) = (600 $\times$ 200)~$\mu$m$^2$. 
All samples are equipped with a metallic Au top-gate, separated from the structure 
by an insulator made of a super-lattice of Si$_3$N$_4$ and SiO$_2$, with a total 
thickness of 110~nm to tune the carrier concentration as a function of the applied gate
voltage $V_g$. Four-probe measurements of longitudinal and transverse electrical
resistances have been carried out using Stanford Research Systems SR830 Lock-In 
amplifiers with low constant voltage excitation. The samples were placed in a 
flow-cryostat in a 33~T Bitter-type magnet. 

\begin{figure}[ht]
\includegraphics[width=0.5\textwidth]{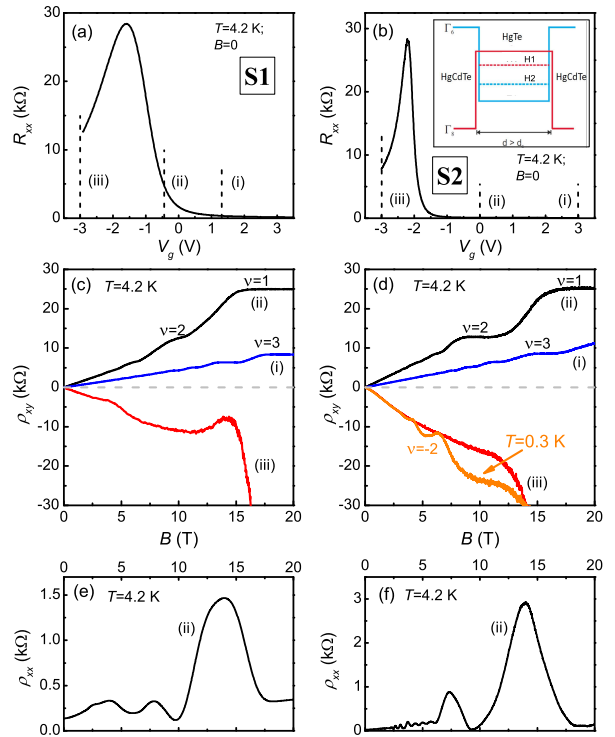}
\caption{
\label{R4p2K} (color online) Magneto-transport in the QSH regime for sample S1 and S2: Gate sweeps at 
$B$=0 for samples S1 (a) and S2 (b). Panels (c) and (d) show the Hall resistivity 
$\rho_{xy} (B)$ for three gate voltage positions indicating the transition 
from electrons, in (i) and (ii), to holes in (iii). The inset in (b) sketches a HgTe 
type-III QW with inverted bands at $T$=4.2~K. When the Fermi energy is in the bulk 
band gap, the $H2$-band is the valence band and the $H1$-band the conduction band.}
\end{figure}

In Fig.~\ref{R4p2K}, we present transport at $T$=4.2~K for sample S1 and sample S2
with a well width of $d$=12~nm. Figures~\ref{R4p2K}(a) and (b)  show the longitudinal 
resistance $R_{xx}$ at $T$=4.2~K as a function of top-gate voltage $V_g$ for samples S1 and S2. 
According to band structure calculations, both QWs are inverted at 4.2~K. The $H1$-band is the 
conduction band and the $H2$-band is the valence band, see sketch of the QW in the 
inset of Fig.~\ref{R4p2K}(b). The finite maximum in $R_{xx}$ is an indication for 
the QSHE \cite{4}. Its value is higher than $h/2e^2$ which can be explained by 
inelastic scattering in large samples \cite{7,8}. This resistance is by an order of 
magnitude smaller than in samples with comparable size and a 20~meV bulk band gap \cite{7} 
(see band structure calculations in Fig.~\ref{BS}(c) and in the supplemental material 
\cite{14}). At $V_g$=0 (ii), both samples are $n$-conducting and sample S1 (S2) has a carrier 
concentration of $n$=3.7$\cdot$10$^{11}~$cm$^{-2}$ ($n$=4.5$\cdot$10$^{11}~$cm$^{-2}$) 
and a mobility of $\mu=5.6\cdot$10$^4$~cm$^2$/Vs ($\mu=4.6\cdot$10$^5$~cm$^2$/Vs). The 
fact that we can indeed tune the Fermi energy through the bulk band 
gap is demonstrated by measuring the Hall resistivity $\rho_{xy}$, shown in 
Figs.~\ref{R4p2K}(c) and (d). Depending on $V_g$, we find a positive 
$\rho_{xy}$ caused by negatively charged electrons in (i) and (ii), and a 
negative $\rho_{xy}$ in (iii) indicating hole transport. At higher magnetic fields, 
we observe the quantum Hall effect for electrons and $\rho_{xy}$ for holes diverges. 
Quantization in $\rho_{xy}$ for holes occurs at lower temperatures, see 
Fig.~\ref{R4p2K}(d) at 0.3~K owing to the higher effective mass for holes \cite{2}.

\section{Magneto-transport at room temperature}

\begin{figure}[ht]
\includegraphics[width=0.5\textwidth]{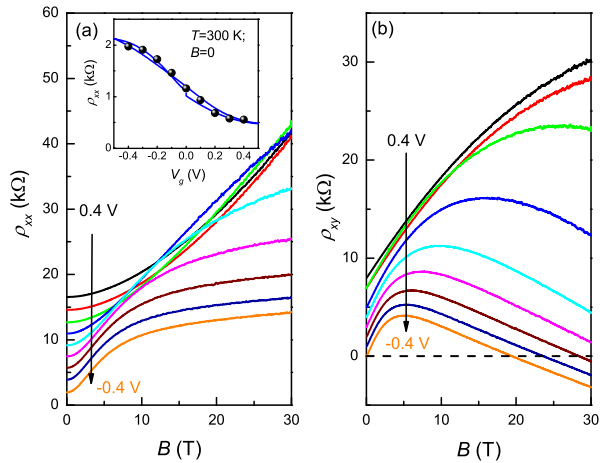}
\caption{
\label{R300K} (color online)
Magneto-transport at room temperature. (a) $\rho_{xx}$ (offset for clarity by 2~k$\Omega$ 
except trace at $V_g$=-0.4~V) and (b) $\rho_{xy}$ (offset for clarity by 1~k$\Omega$ except 
trace at $V_g$=-0.4~V) as a function of $B$ for fixed $V_g$: for high positive $V_g$, $\rho_{xx}$ 
has a quadratic $MR$ accompanied by saturating $\rho_{xy}$, whereas $\rho_{xx}$ exhibits a 
transition from quadratic to almost field-independent MR for negative $V_g$ accompanied 
by a linear decrease in $\rho_{xy}$ at higher fields. Inset: $\rho_{xx}$ as a function of 
$V_g$ at $B$=0 (solid line: continuous gate-sweep; symbols: $\rho_{xx}(B)$ for a fixed $V_g$).}
\end{figure}

We now present magneto-transport at room temperature. At high temperatures we are limited 
to apply a high $|V_{g}|$ due to an increase in the leak current through the
insulator. In Fig.~\ref{R300K}, we show magneto-transport for sample S1 at $T$=300~K. 
Applying a gate voltage $V_g$ at $B$=0, we find that $\rho_{xx}$ increases with 
decreasing $V_g$, see inset of Fig.~\ref{R300K}(a), indicating that we deplete electrons
when decreasing the gate voltage. In Fig.~\ref{R300K}(a) and (b), we illustrate
$\rho_{xx}$ and $\rho_{xy}$ as a function of the magnetic field for various fixed $V_g$. 
For all gate voltages, $\rho_{xx}$ displays a pronounced MR and $\rho_{xy}$ shows a 
strong non-linear behavior. For positive $V_g$, $\rho_{xx}$ increases
quadratically as a function of $B$ but with decreasing $V_g$, we find that $\rho_{xx}$
deviates from the quadratic behavior and exhibits a small $MR$ seemingly saturating 
in higher magnetic fields.

Both observations point towards a system where electrons 
and holes coexist. Notably, the slope of $\rho_{xy}$ is first positive, indicating a 
dominant contribution  of mobile electrons. With increasing magnetic field, the slope 
becomes negative due to holes with a higher concentration and lower mobility. 

We can extract quantitative information on the charge carrier properties using a 
semi-classical Drude-model with field-independent electron and hole concentrations 
and mobilities where we sum up the individual contributions of both electrons and holes 
to the conductivity tensor 
$\widehat{\sigma}$
\begin{equation}
\begin{array}[b]{r}
  \sigma_{xx}=\frac{ne\mu_e}{(1+\mu_e^2B^2)}+\frac{pe\mu_p}{(1+\mu_p^2B^2)}; \\
  \sigma_{xy}=\frac{ne\mu_e^2B}{(1+\mu_e^2B^2)}-\frac{pe\mu_p^2B}{(1+\mu_p^2B^2)}.
\end{array}
	\label{eq1}
\end{equation}
and perform a fit to the experimentally measured resistivity tensor 
$\widehat{\rho}=\widehat{\sigma}^{-1}$.

\begin{figure}[ht]
\includegraphics[width=0.5\textwidth]{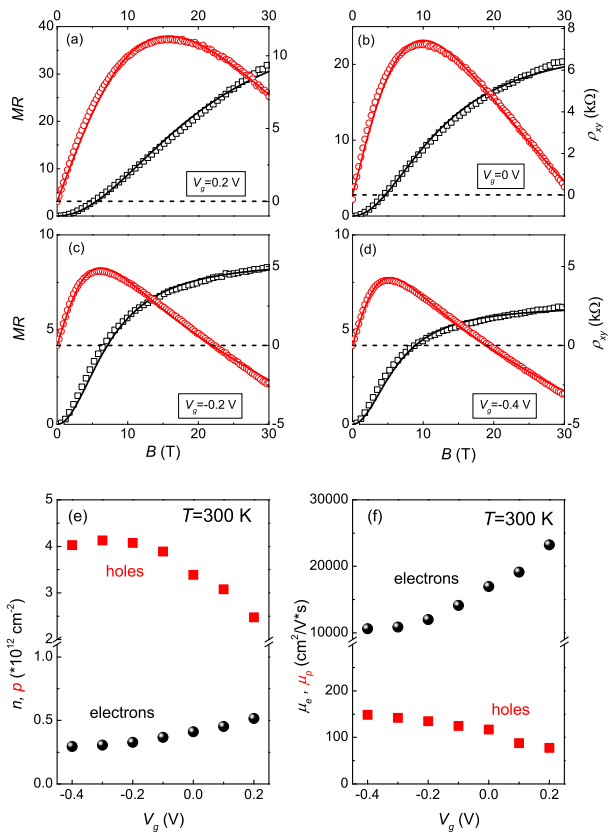}
\caption{
\label{Fits300K} (color online)
Analysis of room-temperature magneto-transport with a two-carrier 
Drude model (symbols represent the experimental data, solid lines fits to the model): 
(a)-(d) $MR$ and $\rho_{xy}$ up to $B$=30~T for several chosen gate voltages. (e) 
Charge carrier concentrations and (f) charge carrier mobilities for electrons and holes 
at different $V_g$ as extracted from the fits. With decreasing gate voltage, the electron 
(hole) concentration and mobility decreases (increases).}
\end{figure}

The results of the simultaneous fitting of the relative magneto-resistance,
defined as $MR=[\rho_{xx}(B)-\rho_{xx}(0)]/\rho_{xx}(0)$, and the Hall resistance
$\rho_{xy} (B)$ at four chosen fixed gate voltages are shown as the solid lines 
in Fig.~\ref{Fits300K}(a-d). The gate dependences of the electron and hole 
concentrations $n$ and $p$, and the mobilities $\mu_e$ and $\mu_p$ extracted 
from the fits are illustrated in Fig.~\ref{Fits300K}(e) and Fig.~\ref{Fits300K}(f).
Therefore, transport is governed by an electron band with very mobile carriers, 
i.e.~carriers with a small effective mass, coexisting with a hole band with a large 
amount of charge carriers with a large effective mass and low mobility.

\section{Temperature-dependent band structure calculations}

So far, we have presented that our system is a 2D TI at low-temperature 
whereas magneto-transport can be described within a classical two-carrier model 
for one electron and one hole band at room temperature as expected for a conventional
semiconductor if the thermal energy $k_B T$ is larger than the band gap $E_g$. 
This transition can be elucidated by performing temperature-dependent
bulk band structure calculations of the QW structures shown for
sample S1 in Fig.~\ref{BS}. Additional BS calculations for sample S2 are 
illustrated in the supplemental material \cite{14}.
Our calculations are based on an eight-band \textbf{k$\cdot$p} model in the envelope 
function approach \cite{15}. The \textbf{k$\cdot$p} model takes into 
account the temperature dependence of all relevant parameters, in particular the 
change in the lattice constants of Hg$_{1-x}$Cd$_x$Te and the elastic constants 
$C_{11}$, $C_{12}$ (bulk modulus) and $C_{44}$ with temperature \cite{16,17}. 
The elastic constants $C_{ij}$ increase by a few $\%$ with decreasing $T$ but 
the ratios which enter the calculations remain constant. In Figs.~\ref{BS}(a-d), 
we plot $E(k)$ for the 12~nm thick QW at different temperatures used in our experiment.
At 300~K, the gap between the conduction band $E1$ and the valence band $H1$ is 
$E_{g}\simeq$26~meV. With decreasing temperature, the band-gap $E_g$ considerably 
decreases and the 12~nm HgTe QW becomes a zero-gap SC at $T$=223~K \cite{10}, 
see Fig.~\ref{BS}(b). In this temperature range $E_g\leq k_B T$, and transport is 
still governed by thermally activated charge carriers within the $E1$ and $H1$ bands. 
Decreasing the temperature further, the BS becomes inverted and the system has an indirect
bulk band gap. Thus, the BS calculations demonstrate a transition from a conventional 
semiconductor with normal band ordering to a 2D TI. At $T$=100~K, see Fig.~\ref{BS}(c), 
transport is still dominated by thermally activated charge carriers but now in the $H1$ 
conduction and the $H2$ valence band since $E_g\leq k_B T$. At low temperatures, 
see Fig.~\ref{BS}(d) for $T$=4.2~K, $E_g>k_B T$ and thermal activation becomes negligible
and we would observe an infinite resistance in the bulk band gap for a perfectly 
homogeneous gate potential if our system was not a 2D TI. 

\begin{figure}[ht]
\includegraphics[width=0.5\textwidth]{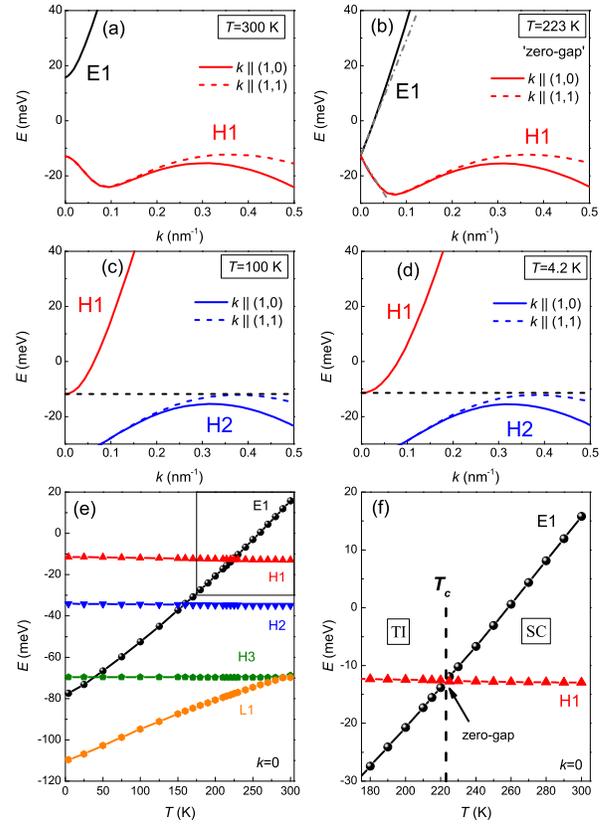}
\caption{
\label{BS} (color online)
Band structure calculations for sample S1 for $k||$(1,0) and $k||$(1,1):
(a)-(d) $E(k)$ for (a) at 300~K with a normal BS ($E1>H1$),
(b) at 223~K where the system is a zero-gap semiconductor (dashed-dotted lines show
$E\propto k$ for small $k$ where the $E1$ and $H1$ bands touch each other at $k$=0),
(c) at 100~K and (d) at 4.2~K with an inverted
BS. We find an indirect bulk band-gap at 4.2~K (Note that the curvature
of the $H1$-band is larger at 100~K compared to 4.2~K).(e) Temperature dependence of the
electron-like $E1$, the heavy-hole like $H1$, $H2$
and $H3$ and the light-hole-like $L1$ subband as a function of temperature
at $k$=0. The $E1$ and $L1$ band decrease with decreasing $T$.
(f) Crossing of $E1$ and $H1$ at $T_{c}$: for $T>$223~K, the system is a normal
semiconductor. For $T<$223~K, we have a QW with an inverted BS and thus, a 2D TI.}
\end{figure}

In Fig.~\ref{BS}(e), we plot an overview of the temperature dependence of the
electron band $E1$, the heavy-hole bands $H1$, $H2$ and $H3$ and the light hole band
$L1$ at the $\Gamma$-point ($k$=0). As can be seen from the calculation, the system 
undergoes a transition from a conventional semiconductor to a 2D TI due to the decrease
of the $E1$-band in energy, highlighted in Fig.~\ref{BS}(f) where the temperature 
dependence of $E1$ and $H1$ is illustrated. For $T>$223~K, our system is a conventional 
semiconductor with the conduction band $E1$ and the valence band $H1$ with a direct band gap 
[indirect band gap] for $k||$(1,0) [$k||$(1,1)].

\section{Regime of linear magneto-resistance}

\begin{figure}[ht]
\includegraphics[width=0.5\textwidth]{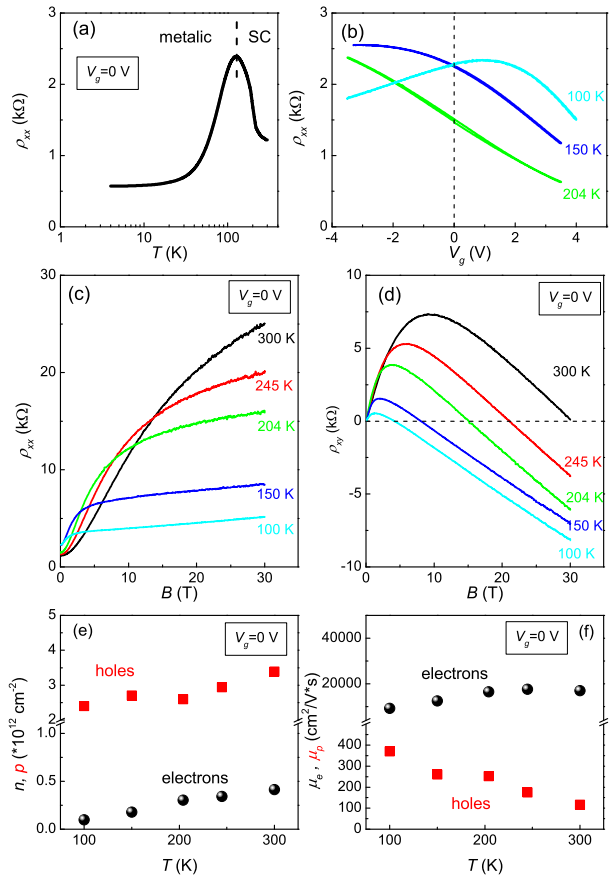}
\caption{
\label{RT} (color online)
Temperature-dependent magneto-transport: (a) $\rho_{xx}(T)$ at
$V_g$=0 shows a maximum around 130~K, then decreases and saturates with decreasing
temperature. (b) $\rho_{xx}$ as a function of $V_g$ at different temperatures. 
For $T$=100~K, $R_{xx}$ exhibits a maximum at $V_g$=1~V which we identify as the CNP.
(c) $\rho_{xx}$ and (d) Hall resistivity as a function of $B$ at $V_g$=0 for 300, 245,
204, 150 and 100~K. For 150 and 100~K, $\rho_{xx}$ exhibits LMR in a wide range of 
magnetic field. (e) Carrier concentrations and mobilities for electrons and holes as 
a function of temperature. The dashed lines (dashed-dotted lines) mark the border 
between two-carrier transport and one-carrier $n$-conduction for $k||$(1,0)
when the thermal energy is smaller than the bulk band gap extracted from
BS calculations.}
\end{figure}

Let us now draw our attention to the intermediate temperature regime where, according
to the presented BS calculation, the band order is inverted but transport is still
dominated by thermally activated charge carriers. In 
Fig.~\ref{RT}(a), we plot $\rho_{xx}(T)$ at $V_g=$0 and observe that $\rho_{xx}$ first 
increases with decreasing temperature, which is characteristic for a semiconductor 
with thermally activated carriers. Around 130~K, $\rho_{xx}$ displays a maximum and then 
starts to decrease with further decreasing temperature before saturating 
for $T\leq$ 10~K. This behavior is characteristic for a metallic system with 
a constant carrier concentration and an increasing mobility with decreasing 
temperature. In Fig.~\ref{RT}(b), we plot the resistivity $\rho_{xx}$ as a function 
of $V_g$ at 100, 150 and 204~K. For $T>100$~K, $\rho_{xx}$ decreases 
monotonously with increasing $V_g$. For $T=100$~K, we observe a maximum in $\rho_{xx}$ 
which we refer to as the region of charge neutrality.

\begin{figure}[ht]
\includegraphics[width=0.5\textwidth]{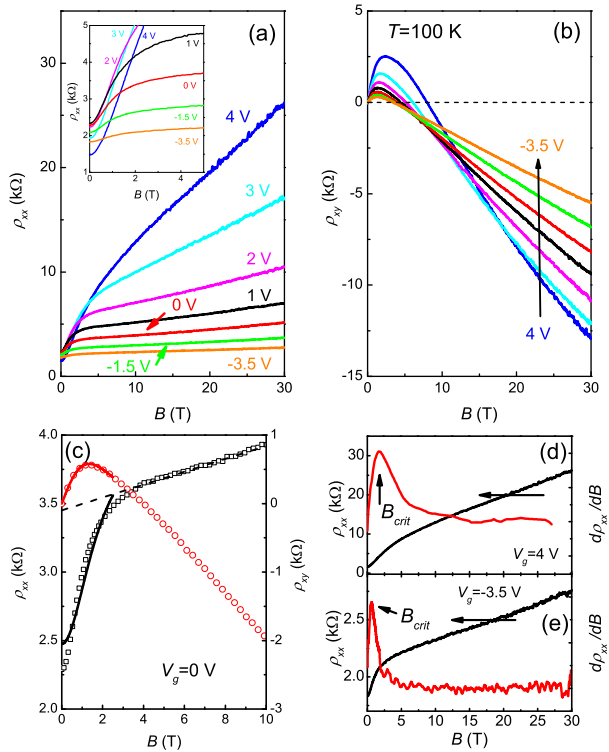}
\caption{
\label{LMR} (color online)
Magneto-transport at $T$=100~K: (a) $\rho_{xx}$ (inset: low-field behavior of 
$\rho_{xx}$) and (b) $\rho_{xy}$ as a function of $B$ 
for different gate voltages. For all $V_g$, $\rho_{xx}$ exhibits
strong LMR. (c) Applied two-carrier fit model (solid lines),
measured $\rho_{xx}$ and $\rho_{xy}$ (open symbols and circles) at $V_g$=0.
(d), (e) The onset of LMR shifts to lower $B$ with decreasing $V_g$ where both
$\rho_{xx}$ and $d\rho_{xx}/dB$ are plotted as a function of $B$ for $V_g$=4~V
and $V_g$=-3.5~V, respectively. $B_{crit}$ marks the transition from quadratic $MR$ to
$LMR$.}
\end{figure}

A particular feature of this intermediate-temperature regime is the emergence of a
strong LMR that develops from a classical quadratic low-field MR. In Figs.~\ref{RT}(c) 
and (d) we plot $\rho_{xx}$ and $\rho_{xy}$ at $V_g$=0 as a function of $B$ for several temperatures. 
In the temperature range presented in Fig. 5(c), $E_g \leq k_B T$ at $V_g$=0 and 
magneto-transport is governed by bulk electrons and holes.
For high temperatures ($T > 200$~K), we observe a quadratic MR that 
can be perfectly modeled by a two-carrier Drude model 
(see also Fig.~\ref{Fits300K}). For lower temperatures, however, we find two 
different regimes in $\rho_{xx}(B)$: a quadratic MR at low magnetic fields, 
as expected from the classical two-carrier Drude model, and a LMR at high $B$. 
Furthermore, the onset of LMR shifts continuously to lower $B$ with decreasing temperature. 
Interestingly, at $T$=150 and 100~K, we observe a wide range of LMR, e.g. from 8 to 30~T at
150~K and 3 to 30~T at 100~K, respectively. The experimentally observed MR can not be
described by the classical two-carrier model though the corresponding $\rho_{xy}$
traces indicate that both electrons and holes still contribute to the 
transport. We estimate the electron concentrations from the linear increase of $\rho_{xy}$ 
at low $B$ and the hole concentration from the slope 
of $\rho_{xy}$ at high $B$ and the results are shown in Fig.~\ref{RT}(e). When we decrease the 
temperature from 300 to 100~K, $n$ and $p$ decrease by approximately a factor of two and 1.5, 
respectively, which can be again be explained by the decrease of thermally exited 
carriers. We can also estimate the carrier mobilities, by limiting the two-carrier 
model to the classical regime at low magnetic fields. An example of a two-carrier 
fit is shown in Fig.~\ref{LMR}(c) for $B\leq$2.5~T (solid lines) at 100~K.

As an example for the peculiar MR below $T_c$ when the band structure is inverted, we
plot $\rho_{xx}$ and $\rho_{xy}$ in Figs.~\ref{LMR}(a) and (b) as a function of the 
magnetic field up to 30~T at different $V_g$, respectively. For $V_g\geq$1~V, $\rho_{xx}$
exhibits a strong positive MR at low fields that evolves into a strong linear MR
with increasing $B$. For $V_g\leq$1~V, the MR is still positive at low $B$ and becomes
linear as the field is increased, see also inset of Fig.~\ref{LMR}(a), and the onset 
of $LMR$ shifts to lower magnetic field with decreasing $V_g$. The linear dependence 
of $\rho_{xx}$ and its onset can be clearly illustrated in the first-order derivative 
$d\rho_{xx}/dB$, as plotted in Fig.~\ref{LMR}(d) and (e) as a function of the magnetic 
field. We define a critical field $B_{crit}$ as the magnetic field corresponding to the maximum 
in $d\rho_{xx}/dB$, which marks the deviation from a squared dependence in the 
two-carrier model for low $B$. For all $V_g$, $\rho_{xy}$ is first positive due to mobile
electrons and becomes negative with increasing $B$ due to the presence of holes.  

\section{Discussion}

From the above bulk BS calculations we see that temperature induces a transition 
from a normal state to a topologically non-trivial state in HgTe QWs. With decreasing 
temperature ($T>T_{c}$), the gap closes, see Fig.~\ref{BS}(e). The conduction band 
exhibits a significant dependence on $k$, yielding a small effective mass 
0.015~m$_e<m^*<$0.04~m$_e$ \cite{2,15,16} for our QW. In contrast, the valence band 
is more flat pointing to a much higher effective mass ($m^*\simeq$0.2~$m_e$ \cite{16}). 
As demonstrated in Fig.~\ref{BS}(f), the transition from normal to inverted band order 
occurs at $T_c$=223~K. Since the thermal energy is larger than $E_g$, 
magneto-transport is dominated by bulk electrons and holes due to the small bulk band 
gap above and below $T_c$. Thus, this transition does not occur abruptly in magneto-transport 
as demonstrated by our experimental data in Figs.~\ref{RT} and~\ref{LMR}.
However, as presented in Fig.~\ref{LMR}(c), the classical two-carrier model fails to describe 
the observed MR for $B>B_{crit}$. Moreover, the peculiar MR occurs if electrons and 
holes with a considerable difference in carrier concentration and mobility coexist. 
The fact that our bulk band structure is inverted for $T<$223~K implies that 
transport can also take place in helical edge states with a linear dispersion 
in the bulk band gap \cite{4,5,6,7}. 

The observation of LMR has been reported in various systems such as bulk 
narrow-band gap semiconductors \cite{18,19} and semi-metals \cite{20} as well as
recently in TIs \cite{21}. In fact, the occurrence of a strong LMR has been ascribed 
to surface states in three-dimensional TIs \cite{22}. In a 2D TI (HgTe QW), LMR has been 
also found at low magnetic field and low temperature when the chemical potential moves 
through the bulk gap \cite{23}. In contrast, since $k_{B}T>E_g$, LMR in our system is 
governed by mobile bulk electrons with low density and less mobile holes with high 
carrier concentration and helical edge states. From $V_g$-dependent 
measurements we found that the onset $B_{crit}$ of LMR shifts to lower $B$ with increasing 
carrier concentration of holes. 

Theoretical models have also addressed the appearance of LMR \cite{24,25}. The classical 
percolation model by Parish and Littlewood \cite{24} for a non-saturating 
LMR due to distorted current paths caused by disorder-induced inhomogeneities in the
electron mobility cannot be applied to our system since our MBE grown samples do not show 
strong fluctuations in $\mu$, and, it does not explain the transition from classical MR 
to LMR with increasing magnetic field. The quantum model, that has been proposed by
Abrikosov \cite{25} is valid for systems with a gapless linear dispersion spectrum
when only the lowest Landau level (LL) remains occupied. Moreover, the energy difference
between the lowest LL $E_0$ and the first LL $E_1$ should be much larger than
$E_F$ and $k_{B}T$. We reach the quantum limit for one type of charge carriers, 
e.g. at $V_g$=0 for $T$=100~K, since $E_{1}-E_{0}>E_F>k_{B}T$, however, the LMR 
in our 2D system occurs in the presence of two types of charge  carriers in the bulk and 
charge carriers in the helical edge states in contrast to the three-dimensional model 
for one type of charge carrier proposed by Abrikosov \cite{25}.

Recently, MR has been theoretically investigated in two-component systems such as
narrow-band semiconductors or semi-metals at high temperatures \cite{26}. For equal carrier 
concentrations of electrons and holes, a non-saturating LMR has been predicted to occur 
in finite size at charge neutrality due to the interplay between bulk and edge contributions.
At room temperature, our data in Fig.~\ref{R300K} shows qualitatively the expected behavior
for $\rho_{xx}$ and $\rho_{xy}$ as proposed and illustrated in Ref.~\cite{26} for broken
electron-hole symmetry. Yet we have shown that both $\rho_{xx}$ and $\rho_{xy}$ can also
be explained within the classical two-carrier model without any contribution due to a 
quasi-particle density that develops near the sample edges. A satisfactory theoretical 
explanation of the origin of LMR for $T<T_{c}$ and $B>B_{crit}$, that also addresses the 
role of helical edge states at high temperature remains open and is certainly challenging 
for theoretical models in the future.

\section{Conclusion}

We have demonstrated in bulk band structure calculations on HgTe QWs that temperature
induces a transition from a semiconductor at room temperature to a TI at low temperature. Experimentally, 
we can distinguish between three regimes in magneto-transport: (i) transport 
of coexisting electrons and holes that can be described within a classical two-carrier model at room 
temperature, (ii) the appearance of a strong LMR for $B>B_{crit}$ and $T<T_{c}$ where
still electrons and holes coexist and (iii) the regime of quantized transport 
($\hbar\omega_{c}>k_{B}T$) at low temperature where we are also in the regime of the
QSHE. We note that apart from inverted HgTe QWs, the only other system known to be a 2D TI 
is the InAs/GaSb hybrid system \cite{27} that has been investigated at low temperature \cite{8,9,28}. 
Temperature-dependent magneto-transport experiments could demonstrate whether the MR effects are unique
in inverted HgTe QWs due to their bulk band structure or are a fundamental property of
2D topological insulators.

\subsection{Acknowledgments}

This work has been performed at the HFML-RU/FOM member of the European Magnetic Field 
Laboratory (EMFL) and has been supported by EuroMagNET II under EU Contract 
No. 228043, by the DARPA MESO project through the contract number N66001-11-1-4105, 
by the German Research Foundation (DFG grant HA5893/4-1 within SPP 1666, the Leibniz 
Program and DFG-JST joint research project ´Topological Electronics´) and the EU 
ERC-AG program (Project 3-TOP). S.W. is financially supported by a VENI grant of the 
Nederlandse Organisatie voor Wetenschappelijk Onderzoek (NWO).

\end{document}